\documentclass[runningheads,a4paper]{llncs}

\usepackage{amsmath}
\usepackage{fixltx2e}
\usepackage{verbatim}

\usepackage{color}

\usepackage{subfigure}
\usepackage[T1, OT1]{fontenc}
\DeclareTextSymbolDefault{\dh}{T1}

\usepackage{listings}

\usepackage{multirow}

\usepackage{amssymb}
\usepackage{graphicx}
\usepackage[font={normalsize},labelfont=bf]{caption}
\usepackage{varwidth}
\usepackage[hyphens]{url}
\usepackage{tcolorbox}
\usepackage{fixltx2e}
\usepackage{array}
\usepackage{booktabs}

\usepackage{xcolor}
\usepackage[
breaklinks=true,
colorlinks=true,
urlcolor=blue]{hyperref}

\makeatletter
\g@addto@macro{\UrlBreaks}{\UrlOrds}
\makeatother
\PassOptionsToPackage{hyphens}{url}
\usepackage{fancyvrb}
\fvset{frame=single,samepage=true,fontfamily=courier,numbers=left,numbersep=2pt,baselinestretch=0.50,fontsize=\small}

\urldef{\mailsa}\path|{maturban, mln, mweigle}@cs.odu.edu|
\urldef{\mailsb}\path|mklein@lanl.gov|
\urldef{\mailsc}\path|herbert.van.de.sompel @dans.knaw.nl|

\newcommand{\keywords}[1]{\par\addvspace\baselineskip
\noindent\keywordname\enspace\ignorespaces#1}

\usepackage{lipsum}

\usepackage{tikz}
\usetikzlibrary{backgrounds}

\usepackage{listings}
\usepackage{xcolor}

\usepackage{courier}

\usepackage{changepage}

\usepackage{rotating}
      \usepackage{multirow}
        \usepackage{bigstrut}
    \usepackage{booktabs}

\lstdefinestyle{base}{
  emptylines=1,
  breaklines=true,
  basicstyle=\footnotesize\ttfamily,
  moredelim=**[is][\color{red}]{@}{@},
  moredelim=**[is][\color{blue}]{$}{$},
}

\begin{document}

\mainmatter  

\title{Collecting 16K archived web pages\\ from 17 public web archives}

\titlerunning{Collecting 16K archived web pages\\ from 17 public web archives}

%
%
\author{Mohamed Aturban$^1$ \and Michael L. Nelson$^1$ \and Michele C. Weigle$^1$ \and Martin Klein$^2$ \and Herbert Van de Sompel$^3$}
\authorrunning{M. Aturban \and M. L. Nelson \and M. C. Weigle \and M. Klein \and H. Van de Sompel}

\institute{$^1$ Dept of Computer Science, Old Dominion University, Norfolk, VA (USA)
\mailsa\\
\vspace{.7mm}
$^2$ Research Library, Los Alamos National Laboratory, Los Alamos, NM (USA)\\
\mailsb\\
\vspace{.7mm}
$^3$ Data Archiving and Networked Services (Netherlands)\\
\mailsc\\
}

%
%

\toctitle{Discovering Mementos in Web Archives}
\tocauthor{Discovering Mementos in Web Archives}
\maketitle

\setcounter{secnumdepth}{3}

\begin{abstract}
We document the creation of a data set of 16,627 archived web pages, or mementos, of 3,698 unique live web URIs (Uniform Resource Identifiers) from 17 public web archives. We used four different methods to collect the dataset. First, we used the Los Alamos National Laboratory (LANL) Memento Aggregator to collect mementos of an initial set of URIs obtained from four sources: (a) the Moz Top 500, (b) the dataset used in our previous study, (c) the HTTP Archive, and (d) the Web Archives for Historical Research group.  
Second, we extracted URIs from the HTML of already collected mementos. These URIs were then used to look up mementos in LANL's aggregator.  
Third, we downloaded web archives' published lists of URIs of both original pages and their associated mementos. 
Fourth, we collected more mementos from archives that support the Memento protocol by requesting TimeMaps directly from archives, not through the Memento aggregator. Finally, we downsampled the collected mementos to 16,627 due to our constraints of a maximum of 1,600 mementos  per archive and being able to download all mementos from each archive in less than 40 hours.

\end{abstract}

\section{Introduction}
%
%
Even though web archives hold billions of archived web pages \cite{iaurilsnew}, or mementos, obtaining a sample of mementos can be difficult. We describe the steps we took to create a data set of 16,627 mementos of 3,698 unique live web URIs (Uniform Resource Identifiers) from 17 public web archives. We use this collection in our study of identifying changes and transformations in the content of mementos over time (our preliminary work can be found in  \cite{aturbanjcdl2019fixity,aturbanjcdl2018archivenow,turbocplusplus}).

To obtain a memento, lookup by URI-R\footnote{URI-R identifies an original resource from the live web (as described in Section 2)} is widely supported by most web archives, but this requires a user to know the URI of an original page. For instance, we expect to find mementos of well-known URI-Rs (e.g., \href{https://www.cnn.com}{\texttt{www.cnn.com}}) in massive web archives, such as the Internet Archive (\href{https://web.archive.org}{\texttt{web.archive.org}}), as these archives try to capture the entire web by employing large-scale web crawlers. Other web archives focus on preserving special collections. For instance, the UK Web Archive (\href{http://webarchive.org.uk/ukwa/}{\texttt{webarchive.org.uk/ukwa/}}) was established with the objective of archiving only UK websites (e.g., \href{https://www.parliament.uk/}{\texttt{www.parliament.uk/}}) \cite{bailey2006building}.  Other web archives, such as \href{https://perma.cc}{\texttt{perma.cc}}, \href{http://www.webcitation.org}{\texttt{webcitation.org}}, and \href{https://archive.is}{\texttt{archive.is}}, capture web pages on demand, so they only preserve pages submitted by users, not through crawling the web. Table \ref{tab:archive-appv} shows a list of 17 public web archives:
\begin{itemize}
\item{\textbf{General}: Archives preserve any web page discovered through large-scale web crawlers.}
\item{\textbf{On-demand}: In general, only web pages (URIs) submitted by users are captured, but the archive might also create archived collections or obtain a copy of collections captured by other archives.}
\item{\textbf{National}: Archives preserve a government or country's web content. They might capture web pages with one or more specific Top Level Domain.}
\item{\textbf{Organizational}: Archives preserve web pages that are about specific organizations, such as the European Union.}
\end{itemize}

\begin{table}[ht]
\setlength\tabcolsep{2pt}
\centering
\caption {\bf A set of 17 public web archives.}

\begin{tabular}{l|l|l}
\hline
\textbf{Archive URI}& \textbf{Archive Name} & \textbf{Purpose} \\
\hline

\href{https://swap.stanford.edu}{\texttt{swap.stanford.edu}}  & Stanford Web Archive Portal & General \\ \hline

\shortstack[l]{\href{http://web.archive.org}{\texttt{web.archive.org}}\\~\\~}  & \shortstack[l]{The Internet Archive\\~\\~} & \shortstack[l]{General and \\on-demand}  \\ \hline

\shortstack[l]{\href{http://archive.bibalex.org}{\texttt{archive.bibalex.org}}\\~\\~\\~}  &  \shortstack[l]{\\Bibliotheca Alexandrina's Internet \\ Archive\\~} & \shortstack[l]{ \\National\\~\\~\\~\\~} \\ \hline

\href{https://arquivo.pt}{\texttt{arquivo.pt}}  & The Portuguese Web Archive (PWA) & National \\  \hline

\href{http://www.collectionscanada.gc.ca}{\texttt{collectionscanada.gc.ca}}  & Library and Archives Canada & National \\ \hline

\href{https://www.digar.ee/arhiiv}{\texttt{digar.ee}}  & The Estonian Web Archive & National \\ \hline

\href{http://nationalarchives.gov.uk}{\texttt{nationalarchives.gov.uk}}  & The National Archives & National \\ \hline

\href{https://vefsafn.is}{\texttt{vefsafn.is}}  & The Icelandic Web Archive & National \\ \hline

\href{http://webarchive.loc.gov}{\texttt{webarchive.loc.gov}}   & Library of Congress Web Archives & National \\ \hline

\href{https://webarchive.org.uk}{\texttt{webarchive.org.uk}}  & The UK Web Archive (UKWA) & National \\ \hline

\shortstack[l]{\href{http://webarchive.proni.gov.uk/}{\texttt{webarchive.proni.gov.uk}}\\~\\~}  &  \shortstack[l]{\\Public Record Office of Northern \\ Ireland (PRONI)} & \shortstack[l]{National\\~\\~}  \\ \hline

\shortstack[l]{\href{https://www.webharvest.gov}{\texttt{webharvest.gov}}\\~\\~}   &  \shortstack[l]{\\Congressional \& Federal Government \\ Web Harvests} & \shortstack[l]{National\\~\\~} \\ \hline

\shortstack[l]{\href{http://archive-it.org}{\texttt{archive-it.org}}\\~} & \shortstack[l]{\\Archive-It - Web Archiving Services\\ for Libraries and Archives} & \shortstack[l]{On-demand\\~\\~}\\ \hline

\href{https://archive.is}{\texttt{archive.is}}  & Archive.is & On-demand \\ \hline

\href{https://perma.cc}{\texttt{perma.cc}} & Perma.cc &  On-demand \\ \hline

\href{https://www.webcitation.org}{\texttt{webcitation.org}}  & WebCite & On-demand \\ \hline

\shortstack[l]{\href{http://europarchive.org}{\texttt{europarchive.org}}} & \shortstack[l]{The European Archive}  & \shortstack[l]{\\Organizational}   \\ \hline

\end{tabular}
\label{tab:archive-appv}
\end{table}

\begin{table*}
\centering
\caption {URI-Ms per archive per year. The data is available in CSV format on GitHub at \href{https://github.com/oduwsdl/mementos-fixity/blob/master/urims-per-year.csv}{\texttt{https://github.com/oduwsdl/mementos-fixity/blob/master/urims- per-year.csv}}.}
\vspace{1.8mm}
\setlength\tabcolsep{1pt}
 \rotatebox{90}{\begin{tabular}{lrrrrrrrrrrrrrrrrrrrrrrr}
\textbf{Archive} & \textbf{URI-Ms}&\textbf{1996}&\textbf{97}&\textbf{98}&\textbf{99}&\textbf{00}&\textbf{01}&\textbf{02}&\textbf{03}&\textbf{04}&\textbf{05}&\textbf{06}&\textbf{07}&\textbf{08}&\textbf{09}&\textbf{10}&\textbf{11}&\textbf{12}&\textbf{13}&\textbf{14}&\textbf{15}&\textbf{16}&\textbf{17}\\ \hline
\href{http://webarchive.loc.gov}{\texttt{webarchive.loc.gov}}    &    1,594 &   - &   1 &   1 &   1  &  4  & 100 & 100 & 100  & 99 & 100 & 100 & 100 & 100  &  98  &  99  &  99  &  99  &  98  &  98 &   99 &   98 &   - \\
\href{https://vefsafn.is}{\texttt{vefsafn.is}}       &        1,589  &  6 &   8 &  10  & 11 &  11 &  13 &  13 &  14 &  42 &  46 &  74  & 71  & 70 &   85 &  102 &  116 &  140 &  153 &  152 &  152  & 150 &  150 \\
\href{https://www.webcitation.org}{\texttt{webcitation.org}}     &     1,585 &   -  &  -  &  - &   -  &  - &   -  &  -  &  -  &  - &  28 &  89 &  85 &  70 &  119 &  156 &  156 &  157 &  156 &  155  & 130 &  127 & 157 \\
\href{https://arquivo.pt}{\texttt{arquivo.pt}}     &         1,569 &  30 &  14  & 14 &  15 &  15 &   - &   - &   - &   - &   1 &   1 &   - & 163  & 167  & 166 &  163 &  162 &  167 &  165 &  164  & 162 &   - \\
\href{http://web.archive.org}{\texttt{web.archive.org}}  &        1,566  & 73 &  73 &  73 &  69   & 71 &  71  & 72 &  73 &  72 &  73 &  72 &  72 &  72 &   72 &   70 &   69  &  69 &   67 &    70 &   71 &   72 &  70 \\
\href{https://archive.is}{\texttt{archive.is}}             &  1,396  & 11 &  10 &   9 &  12 &  10  & 12 &  14 &  13 &  18 &  14 &  20 &  33 &  25  &  29 &   28  &  59  &  12 &  214  & 214 &  214 &  213 & 212 \\
\href{http://archive-it.org}{\texttt{archive-it.org}} &  1,383  & 17 &  15 &   2 &   1 &   3  &  1 &  1 &   - &   1 &  51 & 109 & 107 & 108  & 105 &  109 &  107 &  106 &  109 &  107  & 107 &  109 & 108 \\
\href{https://swap.stanford.edu}{\texttt{swap.stanford.edu}}      &  1,222  &  - &   -  &  -  &  -  &  -  &  -  &  - &   - &   -  &  -  &  -  & 21 &  77 &  185  & 166 &  119 &  135 &  164 &  180 &  140  &  21 &  14 \\
\href{http://nationalarchives.gov.uk}{\texttt{nationalarchives.gov.uk}} &    994 &   - &   - &   - &   -  &  - &   - &   1  &  2 &  25 &  12 &  50 &  40 &  97 &  117 &  106  & 110 &  104  &  94 &   83  &  59 &   54  & 40 \\
\href{http://europarchive.org}{\texttt{europarchive.org}}       &    979  &  -  &  -  &  -  &  -   & -  &  -  &  -  &  -  &  -   & -  &  -  &  -  &  -  &   -  &   - &  120 &  219  &  72  & 172 &  146 &  213 &  37 \\
\href{https://www.webharvest.gov}{\texttt{webharvest.gov}}          &   712  &  -  &  -  &  -  &  -  &  -   & - &   - &   - & 128 &   - & 126  &  -  & 91  &   -  & 129 &    2 &  127 &   59 &   38 &   12  &   - &   - \\
\href{https://www.digar.ee/arhiiv}{\texttt{digar.ee}}                 &  488  &  -  &  -  &  -  &  -  &  -   & - &   -  &  - &   - &   - &   -  &  - &   - &    -  &  36 &   95 &   69 &   89 &   69 &   74 &   56 &   - \\
\href{http://webarchive.proni.gov.uk/}{\texttt{webarchive.proni.gov.uk}}              & 469  &  -  &  -  &  -  &  - &   -   & -   & -  &  - &   - &   -  &  -  &  -  &  -  &   - &   17 &   94 &   19 &   75 &   75 &   78 &   59 &  52 \\
\href{http://www.collectionscanada.gc.ca}{\texttt{collectionscanada.gc.ca}}    & 351 &   - &   - &   - &   - &    - &   -  &  - &   - &   -  & 40 & 173 & 138 &   -  &   - &    - &    - &   -  &    - &    - &    - &    - &   - \\
\href{https://webarchive.org.uk}{\texttt{webarchive.org.uk}}       &   349   & -  &  -   & -  &  - &   -  &  -  &  - &   - &   - &   6 &   9 &  10 &  31 &   34 &   31 &   34 &   34 &   30 &    34 &   29  &  34 &  33 \\
\href{http://archive.bibalex.org}{\texttt{archive.bibalex.org}}      &  199  &  -  &  -  &  1  &  - &   -  &  -  &  - &   -  &  -  &  -  &  1 &   - &   -   &  -  &   -  &  99 &   98  &   - &    - &    - &    - &   - \\
\href{https://perma.cc}{\texttt{perma.cc}}        & 182 &   -  &  - &   -  &  - &   -  &  - &   -  &  -  &  - &   - &   -  &  -  &  -  &   - &    - &    -  &   - &    - &   23 &   53 &   53 &  53 \\
\hline
\textbf{Total}                  & \textbf{16,627} & 137 & 121 & 110 & 109 & 114 & 197 & 201 & 202 & 385 & 371 & 824 & 677 & 904 & 1011 & 1215 & 1442 & 1550 & 1547 & 1635 & 1528 & 1421 & 926 \\
\end{tabular}
}
\label{tab:archive-urim-per-archive-per-year}
\end{table*}


Table \ref{tab:archive-urim-per-archive-per-year} shows the actual number of collected mementos, denoted by URI-M\footnote{URI-M identifies an archived version (memento) of an original resource (as described in Section \ref{sec:background})}, per archive and also the distribution of selected mementos through time. We explain in Section \ref{sec:methodology} how we obtained this set of 16,627 mementos (illustrated in Table \ref{tab:archive-urim-per-archive-per-year}). 

During the process of collecting mementos, we obtained many archived pages, but because of the requirements for our target study, we only selected 16,627 mementos. The requirements include: 

\begin{enumerate}
\item{Downloading mementos is a slow operation and since the bottleneck is the archives themselves, parallelization will not help.  We chose a target of completing the download of all mementos from all the archives within 40 hours.  We also planned to do no more than two such downloads per week in order to limit the load on the archives.}
\item{Since we want to study changes in the playback of mementos over time, we chose 200 as the minimum number of URI-Rs per archive.} 
\item{The number of selected mementos from each web archive should not exceed 1,600. This condition should help reducing the difference between large archives and small archives in terms of the number of sampled mementos.} 
\end{enumerate}

The main purpose of this paper is to document how the dataset of mementos was created so it can be reused by other studies. 

\section{Background} \label{sec:background}
In order to automatically collect portions of the web, some web archives employ web crawling software, such as the Internet Archive's Heritrix \cite{sigurdhsson2010incremental,mohr2004introduction}. Having a set of seed URIs placed in a queue, Heritrix will start by fetching web pages identified by those URIs, and each time a web page is downloaded, Heritrix writes the page to a WARC file \cite{kunze2017warc}, extracts any URIs from the page, places those discovered URIs in the queue, and repeats the process. 

The crawling process will result in a set of archived pages. 
To provide access to their archived pages, many web archives use OpenWayback \cite{openwayback}, the open-source implementation of IA's Wayback Machine, that allows users to query the archive by submitting a URI. OpenWayback will replay the content of any selected archived web page in the browser. One of the main tasks of OpenWayback is to ensure that when replaying a web page from an archive, all resources that are used to construct the page (e.g., images, style sheets, and JavaScript files) should be retrieved from the archive, not from the live web. Thus, at the time of replaying the page, OpenWayback will rewrite all links to those resources to point directly to the archive \cite{tofel2007wayback}. In addition to OpenWayback, PyWb \cite{pywb} is another tool for replaying mementos. It is used by \href{https://perma.cc}{\texttt{perma.cc}}, Webrecorder \cite{webrecorder}, and other archives and services.

Memento \cite{memento:rfc,nelson:memento:tr} is an HTTP protocol extension that uses time as a dimension to access the web by relating the current web resources to their prior states. The Memento protocol is supported by most public web archives including the Internet Archive. The protocol introduces two HTTP headers for content negotiation. First, \texttt{Accept-Datetime} is an HTTP Request header through which a client can request a prior state of a web resource by providing the preferred datetime, for example, 
\\\\
\centerline{\texttt{Accept-Datetime: Mon, 09 Jan 2017 11:21:57 GMT}}.
\\~\\
Second, the \texttt{Memento-Datetime} HTTP Response header is sent by a server to indicate the datetime at which the resource was captured, for instance,
\\\\
\centerline{\texttt{Memento-Datetime: Sun, 08 Jan 2017 09:15:41 GMT}}.
\\
~\\The Memento protocol also defines the following terminology:

\begin{itemize}
\item[--] {URI-R - identifies an original resource from the live web}
\item[--] {URI-M - identifies an archived version (memento) of the original resource at a particular point in time}
\item[--] {URI-T - a resource (TimeMap) that provides a list of mementos (URI-Ms) for a particular original resource (URI-R)}
\item[--] {URI-G - a resource (TimeGate) that supports content negotiation based on datetime to access prior versions of an original resource (URI-R)}
\end{itemize} 

Figure \ref{script:timemap_ex} shows an example of requesting a TimeMap of \href{http://www.cnn.com}{\texttt{www.cnn.com}} from the Internet Archive. This TimeMap has a list of over 227,000 URI-Ms of the original page \href{https://www.cnn.com}{\texttt{www.cnn.com}} captured between June 20, 2000 and March 07, 2019. 

 \begin{figure}[h]
\centering 
\begin{lstlisting}[frame=single,style=base]
% curl  http://web.archive.org/web/$timemap$/link/$http://www.cnn.com$
<http://cnn.com:80/>; rel="original",
<http://web.archive.org/web/timemap/link/http://www.cnn.com>; rel="self"; type="application/link-format"; from="Tue, 20 Jun 2000 18:02:59 GMT",
<http://web.archive.org>; rel="timegate",
<http://web.archive.org/web/20000620180259/http://cnn.com:80/>; rel="$first memento$"; datetime="Tue, 20 Jun 2000 18:02:59 GMT",
<http://web.archive.org/web/20000620180259/http://cnn.com:80/>; rel="$memento$"; datetime="Tue, 20 Jun 2000 18:02:59 GMT",
<http://web.archive.org/web/20000620180259/http://cnn.com:80/>; rel="$memento$"; datetime="Tue, 20 Jun 2000 18:02:59 GMT",

...

<http://web.archive.org/web/20190307012015/https://www.cnn.com/>; rel="$memento$"; datetime="Thu, 07 Mar 2019 01:20:15 GMT",
<http://web.archive.org/web/20190307013003/http://www.cnn.com/>; rel="$memento$"; datetime="Thu, 07 Mar 2019 01:30:03 GMT",
<http://web.archive.org/web/20190307013003/https://www.cnn.com/>; rel="$memento$"; datetime="Thu, 07 Mar 2019 01:30:03 GMT",
\end{lstlisting}
\caption{An example of downloading a TimeMap from the Internet Archive using \texttt{curl}. The TimeMap contains over 227,000 URI-Ms of \href{https://www.cnn.com}{\texttt{www.cnn.com}}.}
\label{script:timemap_ex}
\end{figure}

A Memento aggregator 
can be used to retrieve TimeMaps aggregated from multiple web archives. The Memento Aggregator from Los Alamos National Laboratory (LANL) \cite{bornand2016routing} is one implementation of a Memento aggregator that provides TimeMaps across different web archives both with (a) native support of the Memento protocol and (b) by proxy support of the Memento protocol. MemGator \cite{alam2016memgator,alam2016memgatorgithub} is another implementation of a Memento aggregator and an open source project that provides a variety of customization options, such as allowing users to specify a list of web archives to retrieve TimeMaps from, but it only aggregates TimeMaps from archives that natively support the Memento protocol.

On the  playback of a memento, archives rewrite or transform the original content so that the memento is rendered appropriately in the user's browser. The transformation process includes adding HTML tags to the original content to indicate when the memento was created and retrieved, and rewriting all URI-Rs of embedded resources so they point to the archive, not to the live web. Archives also add banners which provide information about both the memento being viewed and the original page.

In addition to the rewritten content, most archives allow accessing unaltered, raw, archived content (i.e., retrieving the archived version of the original content without any type of transformation by the archive). The most common mechanism to retrieve the raw content, which is supported by different Wayback Machine implementations \cite{openwaybackgithubb,pywb_id}, is by adding \texttt{id\_} after the timestamp in the requested URI-M.

\section{Methodology} \label{sec:methodology}

We collected URI-Rs from four different sources. The first 500 URI-Rs were from Moz \cite{moz500},
which provides a list of the top 500 domains on the web. The second set consists of 1,535 URI-Rs from a previous study \cite{brunelle2015not}  about investigating memento damage. The third set contains 6,657,856 URI-Rs that are publicly available in the HTTP Archive \cite{httparchive}.
The final set of URI-Rs (8,774,352) is from the Web Archives for Historical Research group (WAHR) \cite{wahr17}. 

We included the first two sources even though they are relatively small compared to the other two sources because (1) we wanted our final selected set of URI-Rs to have some top/well-known web pages (i.e., URI-Rs from Moz), and (2) the URI-Rs from the study of memento damage contains a mixture of URI-Rs with different path lengths (e.g., \texttt{www.example.com/path/to/file.html}). The main characteristic of URI-Rs that belong to the first and third source (i.e., Moz and HTTP Archive) is that the URI-R consists of a domain name only (e.g., \href{https://www.example.com}{\texttt{www.example.com}}).  
The URI-Rs from WAHR are extracted from tweets about the hashtags \texttt{\#climatemarch}, \texttt{\#MarchForScience}, \texttt{\#porteouverte}, \texttt{\\\#paris}, \texttt{\#Bataclan}, \texttt{\#parisattacks}, \texttt{\#WomensMarch}, and \texttt{\#YMMfire} between December 11, 2015 and May 3, 2017. Table \ref{tab:uri-count} shows the number of collected URI-Rs including the number of URI-Rs by hashtag for the WAHR source. The total number of unique URI-Rs from all four sources is 8,220,606 after removing duplicates.

\begin{table}
\centering
\vspace{-2mm}
\caption{{\bf URI-R source count.}}
\setlength{\tabcolsep}{3pt}
\begin{tabular}{|l|l|l|r|r|}
\hline
\multicolumn{2}{|l|}{\textbf{\shortstack[c]{NAME~\\~\\~\\~}}} & \textbf{\shortstack[c]{Access time~\\~\\~\\~}} & \textbf{\shortstack[c]{URI-Rs{\color{white}llll}\\~\\~\\~}} & \textbf{\shortstack[l]{~\\URI-Rs \\ after removing \\ duplicates} } \\ \hline

\multicolumn{2}{|l|}{MOZ}           &     2017-06-08                 &        500  &    500   \\ \hline

\multicolumn{2}{|l|}{Memento damage}           & 2017-06-08&1,535          &  1,535    \\ \hline

\multicolumn{2}{|l|}{HTTP Archive}          & 2017-04-15 &6,657,856       &   6,657,856       \\ \hline

\multirow{5}{*}{ \shortstack[l]{~\\~\\~\\~\\~\\~\\~\\~\\~\\~\\~\\~\\~\\~\\WAHR} }   & \shortstack[l]{ \#climatemarch\\~}  &  \shortstack[l]{\\ 2017-04-19 \textbf{--} \\ 2017-05-03}             &     \shortstack[l]{175,278\\~}       & \shortstack[l]{41,674\\~}      \\ \cline{2-5} 

                        & \shortstack[l]{\#MarchForScience\\~}   & \shortstack[l]{\\ 2017-04-12 \textbf{--} \\ 2017-04-26} &     \shortstack[l]{299,124\\~}          & \shortstack[l]{90,318\\~}   \\ \cline{2-5} 

                        & \shortstack[l]{\\ \#porteouverte \\ \#paris \\ \#Bataclan \\ \#parisattacks \\ ~}    &        \shortstack[l]{ 2015-12-11\\~\\~\\~\\~\\~}             &      \shortstack[l]{5,561,037\\~\\~\\~\\~\\~}   &  \shortstack[l]{857,490\\~\\~\\~\\~\\~}      \\ \cline{2-5} 

                        & \shortstack[l]{\#WomensMarch\\~}   &      \shortstack[l]{\\ 2017-01-12 \textbf{--} \\ 2017-01-28}               &       \shortstack[l]{2,403,637\\~}     &  \shortstack[l]{526,903\\~}   \\ \cline{2-5} 

                        & \#YMMfire   &          2016-08-20            &       335,276      &  45,327  \\ \hline
\multicolumn{2}{|l|}{\textbf{Total}} &  & \textbf{15,434,243} & \textbf{8,220,606} \\ \hline

\end{tabular}
\label{tab:uri-count}
\vspace{-2mm}
\end{table}


We merged all 8,220,606 unique URI-Rs from the four sources into a single list. The order of how URI-Rs are placed on the list is as follows:

\begin{enumerate}
\item{Moz's URI-Rs were placed on the top of this list followed by URI-Rs from our Memento damage study.}
\item{We repeatedly selected 10 URI-Rs from HTTP Archive and 10 URI-Rs from WAHR, choosing 10 from a different hashtag each round.}
\end{enumerate}

{\raggedleft{}T}he order of URI-Rs in the list is important because we decided to work with a smaller number of URI-Rs for our study. Thus, out of 8,220,606 URI-Rs, we only selected the first 10,000 URI-Rs that fulfill the conditions explained in Section \ref{cond:path1}.

\subsection{Method 1: selecting the first 10,000 URI-Rs from the initial set of 8,220,606 URI-Rs} \label{cond:path1}
URI-Rs must be canonicalized to determine whether or not a URI-R with a particular domain name and file path length has already been selected. We used the canonicalization function 
that is part of PyWb \cite{pywb}. The function indicates that \href{http://www.example.com}{\texttt{http://www.example.com}}, \href{http://www.example.com:80}{\texttt{http://www.example.com:80}}, and \href{http://www.EXAMPLE.com}{\texttt{www.EXAMPLE.com}} are the same, as shown in Figure \ref{script:canon}. The output of this canonicalization function is in Sort-friendly URI Reordering Transform (SURT)  format \cite{surtgithub}.

 \begin{figure}
\centering
\vspace{-2mm}
\begin{lstlisting}[frame=single,style=base]
% python canonicalize.py http://www.example.com
$com,example)/$

% python canonicalize.py http://www.example.com:80
$com,example)/$

% python canonicalize.py www.EXAMPLE.com
$com,example)/$
\end{lstlisting}
\caption{An example showing three different URI-Rs that map to the same URI-R  (in SURT format \cite{surtgithub}) using the canonicalization function from \cite{pywb}}
\vspace{-2mm}
\label{script:canon}
\end{figure}
 
In addition to the canonicalization function, we issued an HTTP HEAD request to discover if two URI-Rs redirect to the same web resource. As Figure \ref{script:head-reguest} shows, sending a HTTP HEAD request to \href{http://www.fb.com}{\texttt{www.fb.com}} and \href{http://facebook.com}{\texttt{facebook.com}} will result in a ``301'' redirect to  \href{https://www.facebook.com/}{\texttt{https://www.facebook.com/}}, which is the URI-R we select, rather than the first two URI-Rs.

 \begin{figure}
\centering
\vspace{-2mm}
\begin{lstlisting}[frame=single,style=base]
% curl -sIL fb.com | egrep -i "(HTTP/|^location:)"
HTTP/1.1 301 Moved Permanently
Location: $https://www.facebook.com/$
HTTP/2 200 

% curl -sIL https://facebook.com | egrep -i "(HTTP/|^location:)"
HTTP/2 301 
location: $https://www.facebook.com/$
HTTP/2 200

\end{lstlisting}
\caption{An example showing two different URI-Rs that redirect to the same URI-R.}
\label{script:head-reguest}
\vspace{-2mm}
\end{figure}

Also, the selected URI-Rs must contain a variety of file path lengths that we group into the following five sets, each of which contains 2,000 URI-Rs with:
\begin{itemize}
\item{$s_0$: Path length of zero
 
- \href{http://www.example.com}{\texttt{www.example.com}}}
\item{$s_1$: Path length of one

- \texttt{www.example.com/file1.html}}
\item{$s_2$: Path length of two 

- \texttt{www.example.com/1/file2.html}}
\item{$s_3$: Path length of three

- \texttt{www.example.com/1/2/file3.html}}
\item{$s_{4+}$: Path length of four or more

- \texttt{www.example.com/1/2/3/file4.html}}
\end{itemize}


{\raggedleft{}T}he final two conditions for selecting the first 10,000 URI-Rs are:

\begin{enumerate}
\item{URI-Rs with the same file path length should not have the same domain name. For example, if \href{http://www.youtube.com/watch?v=cpPG0bKHYKc}{\texttt{www.youtube.com/watch?v=cpPG0bKHYKc}} has already been selected, then \href{http://www.youtube.com/watch?v=hFhiV5X5QM4}{\texttt{www.youtube.com/watch?v=hFhiV5X5QM4}} will not be selected. This may help to collect more unique URI-Rs and vary the content we plan to study.} 
\item{The TimeMaps of selected URI-Rs must contain at least one memento  as our further work is to study any change or transformation in the content of mementos over time.}
\end{enumerate}

To retrieve TimeMaps, we used the LANL Memento Aggregator. Once a TimeMap is downloaded, we reduced the number of mementos in the TimeMap to one memento per year from each archive. TimeMaps returned from LANL's aggregator have more information and metadata than we need for our further study. Therefore, we wrote two Python scripts available on Github\footnote{\href{https://github.com/oduwsdl/mementos-fixity}{\texttt{https://github.com/oduwsdl/mementos-fixity}}}. The script \texttt{timemap.py} extracts only URI-Ms and their \texttt{Memento-Datetime} from the returned TimeMaps, while the second script \texttt{yearly-filter.py} filters TimeMaps by selecting one memento (the first) per year by archive. Figure \ref{script:timemap_complete} shows an example of a TimeMap with 64 mementos of the URI-R \href{http://www.futureofmusic.org/about/positions.cfm}{\texttt{http://www.f\\utureofmusic.org/about/positions.cfm}}, and Figure \ref{script:timemap_filtered} shows the corresponding TimeMap after filtering. It contains only 10 mementos.

\begin{figure}
\centering
\vspace{-2mm}
\begin{lstlisting}[frame=single,style=base]

% python timemap.py http://www.futureofmusic.org/about/positions.cfm > full-timemap.txt 
% cat full-timemap.txt 
20120328211040 http://www.$webcitation.org$/66VfNacdz
20141021161223 http://$archive.is$/20141021161223/http://www.futureofmusic.org/about/positions.cfm
20141021175005 http://$archive.is$/20141021175005/http://www.futureofmusic.org/about/positions.cfm
20141021175817 http://$archive.is$/20141021175817/http://www.futureofmusic.org/about/positions.cfm
20141106145319 http://$archive.is$/20141106145319/http://www.futureofmusic.org/about/positions.cfm
20141106151301 http://$archive.is$/20141106151301/http://www.futureofmusic.org/about/positions.cfm
20070114182707 https://$web.archive.org$/web/20070114182707/http://www.futureofmusic.org:80/about/positions.cfm
... <18 mementos from 2007-2008> ...
20090122061339 https://$web.archive.org$/web/20090122061339/http://futureofmusic.org:80/about/positions.cfm
20090228213737 https://$web.archive.org$/web/20090228213737/http://futureofmusic.org:80/about/positions.cfm
20120607045812 https://$web.archive.org$/web/20120607045812/http://www.futureofmusic.org/about/positions.cfm
20120607045828 https://$web.archive.org$/web/20120607045828/http://futureofmusic.org/about/positions.cfm
20130323010922 https://$web.archive.org$/web/20130323010922/http://www.futureofmusic.org/about/positions.cfm
20130323011136 https://$web.archive.org$/web/20130323011136/http://futureofmusic.org/about/positions.cfm
20131231022915 https://$web.archive.org$/web/20131231022915/http://www.futureofmusic.org/about/positions.cfm
20140819212552 https://$web.archive.org$/web/20140819212552/http://www.futureofmusic.org/about/positions.cfm
20150320143837 https://$web.archive.org$/web/20150320143837/http://www.futureofmusic.org/about/positions.cfm
20160325184708 https://$web.archive.org$/web/20160325184708/http://www.futureofmusic.org/about/positions.cfm
20070114182707 https://$web.archive.org$/web/20070114182707/http://www.futureofmusic.org:80/about/positions.cfm
20070209043456 https://$web.archive.org$/web/20070209043456/http://www.futureofmusic.org:80/about/positions.cfm
... <26 duplicate mementos from web.archive.org> ...
\end{lstlisting}
\caption{The TimeMap of \href{http://www.futureofmusic.org/about/positions.cfm}{\texttt{www.futureofmusic.org/about/positions.cfm}} contains 64 mementos from three different archives: \href{https://web.archive.org}{\texttt{web.archive.org}}, \href{https://archive.is}{\texttt{archive.is}}, and \href{https://www.webcitation.org}{\texttt{webcitation.org}}. 
}
\label{script:timemap_complete}
\vspace{-2mm}
\end{figure}

 \begin{figure}[h]
\centering
\vspace{-2mm}
\begin{lstlisting}[frame=single,style=base]
% cat full-timemap.txt | python yearly-filter.py > yearly-filter.txt 
% cat yearly-filter.txt 
20120328211040 http://www.$webcitation.org$/66VfNacdz
20141021161223 http://$archive.is$/20141021161223/http://www.futureofmusic.org/about/positions.cfm
20070114182707 https://$web.archive.org$/web/20070114182707/http://www.futureofmusic.org:80/about/positions.cfm
20080109053549 https://$web.archive.org$/web/20080109053549/http://www.futureofmusic.org:80/about/positions.cfm#ed
20090122061339 https://$web.archive.org$/web/20090122061339/http://futureofmusic.org:80/about/positions.cfm
20120607045812 https://$web.archive.org$/web/20120607045812/http://www.futureofmusic.org/about/positions.cfm
20130323010922 https://$web.archive.org$/web/20130323010922/http://www.futureofmusic.org/about/positions.cfm
20140819212552 https://$web.archive.org$/web/20140819212552/http://www.futureofmusic.org/about/positions.cfm
20150320143837 https://$web.archive.org$/web/20150320143837/http://www.futureofmusic.org/about/positions.cfm
20160325184708 https://$web.archive.org$/web/20160325184708/http://www.futureofmusic.org/about/positions.cfm
\end{lstlisting}
\vspace{-4mm}
\caption{The TimeMap of \href{https://www.futureofmusic.org/about/positions.cfm}{\texttt{www.futureofmusic.org/about/positions.cfm}} after filtering. It contains only 10 mementos (the first memento per year is selected from each archive). 
}
\label{script:timemap_filtered}
\vspace{-3mm}
\end{figure}

Table \ref{tab:archive-urir-count-per-source-and-path} shows the number of selected URI-Rs per source and path length and Table \ref{tab:archive-urirs-status-code} shows that 13\% of the selected URI-Rs currently have either the HTTP status code 4xx or 5xx. Even though these URI-Rs are no longer live, they are archived.

\begin{table}[h]
\centering
\vspace{-2mm}
\setlength\tabcolsep{7pt}
\caption {\bf The initial collected set of URI-Rs per source by path length (results of Method 1).}
\vspace{2mm}
\begin{tabular}{l|rrrrr|r}
\hline
&  \multicolumn{5}{c|}{\textbf{Path length}} &   \\ \cline{2-6}
\textbf{Source} &  \textbf{s\textsubscript{0}}&  \textbf{s\textsubscript{1}}&  \textbf{s\textsubscript{2}}&  \textbf{s\textsubscript{3}}&  \textbf{s\textsubscript{4+}} & \textbf{Total} \\ \hline

\textbf{MOZ}   &    286   &    17 & 3 & 2 & 1 & \textbf{309}  \\ \hline

\textbf{HTTP Archive}   & 1,581      & 42    & 70  & 2  & 0  & \textbf{1,695}  \\ \hline
\textbf{Memento Damage}   & 114       &  63   & 62  & 42  & 60  & \textbf{341}   \\ \hline
\textbf{\#climatemarch}   & 4       & 74    & 98 & 89  & 99  & \textbf{364}  \\ \hline
\textbf{\#MarchForScience}   & 1       &  162   & 173  & 139  & 243  & \textbf{718}   \\ \hline
\textbf{\shortstack[l]{\\ \#porteouverte \\ \#paris \\ \#Bataclan \\ \#parisattacks} }   &  \shortstack[l]{8\\~\\~\\~\\~\\~}  & \shortstack[l]{758\\~\\~\\~\\~\\~}  & \shortstack[l]{716\\~\\~\\~\\~\\~}  & \shortstack[l]{855\\~\\~\\~\\~\\~} & \shortstack[l]{711\\~\\~\\~\\~\\~} & \shortstack[l]{\textbf{3,048}\\~\\~\\~\\~\\~}  \\ \hline
\textbf{\#WomensMarch}   & 3      & 723     & 734  & 749  & 734  & \textbf{2,943}   \\ \hline
\textbf{\#YMMfire}   & 3      & 161     & 144  & 122  & 152  & \textbf{582}   \\ \hline
\textbf{Total}   & \textbf{2,000}       & \textbf{2,000}     & \textbf{2,000}  & \textbf{2,000}  & \textbf{2,000}  & \textbf{10,000}   \\
\hline
\end{tabular}
\label{tab:archive-urir-count-per-source-and-path}
\vspace{-2mm}
\end{table}

\begin{table}
\centering
\vspace{-2mm}
\setlength\tabcolsep{15pt}
\caption {\bf The final URI-R HTTP status codes of the initial collected set of URI-Rs (results of Method 1).}
\vspace{2mm}
\begin{tabular}{l|rc|r}
\hline
 &  \multicolumn{2}{c|}{\textbf{HTTP status code}} &   \\ \cline{2-3}
\textbf{Path length} &  \textbf{200}&  \textbf{4xx/5xx}& \textbf{Total} \\ \hline

\textbf{s\textsubscript{0}} &  1,870  &       130     & \textbf{2,000}  \\ \hline
\textbf{s\textsubscript{1}} & 1,651  & 349     & \textbf{2,000}  \\ \hline
\textbf{s\textsubscript{2}}  & 1,715  &       285    & \textbf{2,000}  \\ \hline
\textbf{s\textsubscript{3}} &  1,720    &     280    & \textbf{2,000}  \\ \hline
\textbf{s\textsubscript{4+}} & 1,731  & 269     & \textbf{2,000} \\
\hline
\textbf{Total}      &  \textbf{8,687} & \textbf{1,313} & \textbf{10,000} \\
\end{tabular}
\vspace{-2mm}
\label{tab:archive-urirs-status-code}
\end{table}

Table \ref{tab:archive-urim-count-all} (column \textbf{Method 1}) shows the list of 16 web archives from which the mementos of our 10,000 URI-Rs are collected (there is one archive, \href{https://nationalarchives.gov.uk}{\texttt{nationalarc\\hives.gov.uk}}, that has not been counted yet because it has not contributed any mementos). The total number of URI-Rs in the table exceeds 10,000 because a URI-R often has mementos in multiple archives, resulting in some URI-Rs being counted multiple times, but the total number of unique URI-Rs is still 10,000. The total number of URI-Ms in all TimeMaps is 12,988,039. This number drops to 48,199 URI-Ms after applying the one memento per year filter.

\begin{table}
\begin{adjustwidth}{-0.20in}{0in}
\centering
\vspace{-2mm}
\scriptsize
\caption {\bf The four methods used to collect URI-Rs/URI-Ms. The table indicates (shown in bold) that (1) seven archives satisfy the condition of 200 URI-Rs by Method1 (2) five additional archives satisfy the condition of 200 URI-Rs by \textbf{Method 2}, (3) four other archives satisfy the condition by \textbf{Method 3}, and (4) the last archive that satisfies the condition of 200 URI-Rs by \textbf{Method 4}.}

\begin{tabular}{l|rr|rr|rr|rr}
\hline
&  \multicolumn{2}{c|}{\textbf{Method 1}}&  \multicolumn{2}{c|}{\textbf{Method 2}}&  \multicolumn{2}{c|}{\textbf{Method 3}}&  \multicolumn{2}{c}{\textbf{Method 4}} \\

\textbf{Archive} & \textbf{URI-Ms}&\textbf{URI-Rs} & \textbf{URI-Ms}&\textbf{URI-Rs} & \textbf{URI-Ms}&\textbf{URI-Rs}& \textbf{URI-Ms}&\textbf{URI-Rs}\\ \hline

\href{http://web.archive.org}{\texttt{web.archive.org}}   & \textbf{32,139}  &  \textbf{9,790} & 40,258 & 10,353 & 45,155 &    10,924 & 45,155  &  10,924  \\

\href{https://archive.is}{\texttt{archive.is}} & \textbf{2,322} &  \textbf{1,284} & 3,229 & 1,526 & 3,471 & 1,654 & 3,471 & 1,654 \\

\href{http://archive-it.org}{\texttt{archive-it.org}} & \textbf{3,500} & \textbf{804} & 7,986 & 1,355 & 8,994 & 1,593 & 8,994 & 1,593 \\

\href{http://archive.bibalex.org}{\texttt{archive.bibalex.org}}  & \textbf{3,363} &  \textbf{568} & 6,176 & 941 & 7,286 & 1,148 & 7,286 & 1,148 \\

\href{http://webarchive.loc.gov}{\texttt{webarchive.loc.gov}} & \textbf{2,721} & \textbf{418} & 7,122 & 934 & 7,766 & 1,062 & 7,766   &  1,062 \\

\href{https://arquivo.pt}{\texttt{arquivo.pt}} & \textbf{1,410} & \textbf{324} & 3,154 & 758 & 3,430 & 895 & 3,430 &     895 \\

\href{https://www.webcitation.org}{\texttt{webcitation.org}} & \textbf{1,125} & \textbf{472} & 1,858 & 725 & 1,954 & 775 & 1,954     & 775 \\

\href{http://europarchive.org}{\texttt{europarchive.org}} & 407 & 106 & \textbf{911} & \textbf{287} & 992 & 324 & 992  &    324  \\

\href{https://swap.stanford.edu}{\texttt{swap.stanford.edu}} & 609 & 132 & \textbf{1,176} & \textbf{283} & 1,233 & 304 & 1,233 & 304  \\

\href{https://vefsafn.is}{\texttt{vefsafn.is}} & 19 & 7 & \textbf{1,520} & \textbf{246} & 1,715 & 294 & 1,715 & 294 \\

\href{https://www.webharvest.gov}{\texttt{webharvest.gov}} & 84 & 21 &  \textbf{743} & \textbf{227} & 826 & 248 & 826 & 248 \\

\href{https://webarchive.org.uk}{\texttt{webarchive.org.uk}}   & 12  & 5 & 27 & 8 & \textbf{907} & \textbf{228} & 907 & 228 \\

\href{https://www.digar.ee/arhiiv}{\texttt{digar.ee}} & 333 & 129 & \textbf{513} & \textbf{223} & 518 & 228 & 518      & 228  \\

\href{http://webarchive.proni.gov.uk/}{\texttt{webarchive.proni.gov.uk}} & 138  &  48 & 316 & 141 & \textbf{480} & \textbf{213} & 480 & 213  \\

\href{http://nationalarchives.gov.uk}{\texttt{nationalarchives.gov.uk}} & 0 & 0 & 0 & 0 & \textbf{1,011} & \textbf{200} & 1,011 &      200 \\

\href{http://www.collectionscanada.gc.ca}{\texttt{collectionscanada.gc.ca}}   & 8 & 6 & 59 & 50 & \textbf{359} & \textbf{200} & 359 & 200 \\ 

\href{https://perma.cc}{\texttt{perma.cc}}  & 9 & 6 & 101 & 71 & 154 & 111 & \textbf{290} & \textbf{200}  \\

\hline
\textbf{Total} &  \textbf{48,199} & \textbf{14,120}&  \textbf{75,149} & \textbf{18,128}&  \textbf{86,251} & \textbf{20,401}&  \textbf{86,387} & \textbf{20,490}  \\
\end{tabular}
\vspace{-2mm}
\label{tab:archive-urim-count-all}
\end{adjustwidth}
\end{table}


From Table \ref{tab:archive-urim-count-all}, we notice that several archives have a small number of URI-Rs and URI-Ms. Since we want to study the playback fidelity of the web archives, we chose 200 as the minimum number of URI-Rs per archive. After applying Method 1, we used the three methods (Sections 3.2, 3.3, and 3.4) to discover additional mementos from web archives that have fewer than 200 URI-Rs.



\subsection{Method 2: Discovering additional URI-Rs from the HTML of already collected mementos} \label{sec:parsingcollectedmementos}
For each archive that has not satisfied the 200 URI-Rs condition we downloaded the raw content of already collected mementos from the archive and extracted all URI-Rs found in the HTML. Using the LANL Memento Aggregator, we requested the TimeMap of each URI-R that had not already been selected. We applied this method for the following archives:
\begin{enumerate}
\item{\href{https://www.webharvest.gov}{\texttt{webharvest.gov}}}
\item{\href{https://swap.stanford.edu}{\texttt{swap.stanford.edu}}}
\item{\href{https://vefsafn.is}{\texttt{vefsafn.is}}}
\item{\href{https://webarchive.org.uk}{\texttt{webarchive.org.uk}}}
\item{\href{http://webarchive.proni.gov.uk/}{\texttt{webarchive.proni.gov.uk}}}
\item{\href{http://www.collectionscanada.gc.ca}{\texttt{collectionscanada.gc.ca}}}
\item{\href{https://perma.cc}{\texttt{perma.cc}}}
\end{enumerate}

{\raggedleft{}T}hree archives are not included in the list above. The first reason being that Method 2 can not be applied for \href{http://nationalarchives.gov.uk}{\texttt{nationalarchives.gov.uk}} because the archive has not yet provided any mementos. The second is that the archives \href{http://europarchive.org}{\texttt{europarchive.org}} and \href{https://www.digar.ee/arhiiv}{\texttt{digar.ee}} satisfied the condition of 200 URI-Rs after applying Method 2 to \href{https://swap.stanford.edu}{\texttt{swap.stanford.edu}} and \href{https://vefsafn.is}{\texttt{vefsafn.is}}, respectively.

As shown in Table \ref{tab:archive-urim-count-all-method1}, any new discovered URI-Rs/URI-Ms with this method caused the information from all archives to be updated even for archives that already had more than 200 URI-Rs. Figure \ref{script:extrcat-example} shows an example of URI-Rs extracted from the HTML of the memento:
\\\\
\href{http://wayback.vefsafn.is/wayback/20041020191800id_/http://www.w3.org/}{ \texttt{https://wayback.vefsafn.is/wayback/20041020191800id\_/http\\://www.w3.org/}}
\\\\
{\raggedleft{}T}he URI-Rs are extracted from the attribute \texttt{href} in the \texttt{<a>} tags (using the Python script \texttt{extract\_urirs.py}\footnote{\href{https://github.com/oduwsdl/mementos-fixity}{\texttt{https://github.com/oduwsdl/mementos-fixity}}}). We downloaded the TimeMap of the URI-R \href{https://www.inria.fr}{\texttt{www.inria.fr/}} which had not previously been selected. As Figure \ref{script:extrcat-example-timemap} shows, the TimeMap does not only contain mementos from \href{https://vefsafn.is}{\texttt{vefsafn.is}} but also mementos from the eight archives: \href{http://web.archive.org}{\texttt{web.archive.org}}, \href{http://archive.bibalex.org}{\texttt{archive.bibalex.org}}, \href{https://www.webcitation.org}{\texttt{webcitation.org}}, \href{http://webarchive.loc.gov}{\texttt{webarchive.loc.gov}}, \href{http://archive-it.org}{\texttt{archive-it.org}}, \href{https://archive.is}{\texttt{archive. is}}, \href{https://vefsafn.is}{\texttt{vefsafn.is}}, and \href{https://www.digar.ee/arhiiv}{\texttt{digar.ee}}.

\begin{table}
\begin{adjustwidth}{-.1in}{0in}
\setlength{\tabcolsep}{0.05em}
\centering
\scriptsize
\caption {\bf After applying Method 2 to seven archives, five archives satisfy the condition of 200 URI-Rs (shown in bold). Notice that applying this method to one archive may increase the number of URI-Rs in other archives (e.g., applying method 2 for \texttt{vefsafn.is} makes both \texttt{vefsafn.is} and \texttt{digar.ee} satisfy the 200 URI-R condition).}
\begin{tabular}{l|rr|rr|rr|rr}
\hline
& \multicolumn{8}{c}{\textbf{Method 2}} \\ \cline{2-9}

&  \multicolumn{2}{c|} {\href{https://www.webharvest.gov}{\textbf{\shortstack[l]{webharvest.gov\\~\\~\\~\\~\\~}}}}
&  \multicolumn{2}{c|} { \href{https://swap.stanford.edu} {\textbf{\shortstack[l]{swap.stanford\\.edu\\~\\~\\~\\~}}} }
&  \multicolumn{2}{c|}{ \href{https://vefsafn.is} {\textbf{\shortstack[l]{vefsafn.is\\~\\~\\~\\~\\~ \vspace{.5mm} }}} }
&  \multicolumn{2}{c}{
\textbf{
\shortstack[l]{
\href{https://webarchive.org.uk}{webarchive.org}
\\
\href{https://webarchive.org.uk}{.uk}
\vspace{.5mm}  
\\
\href{http://webarchive.proni.gov.uk/}{proni.gov.uk}
\vspace{.5mm}  
\\
\href{http://www.collectionscanada.gc.ca}{collectionscanada}
\\
\href{http://www.collectionscanada.gc.ca}{.gc.ca} 
\vspace{.5mm} \\ 
\href{https://perma.cc}{perma.cc}
\vspace{.5mm} 
\\~
}
}
}  \\ \cline{2-9}

\textbf{Archive} & \textbf{URI-Ms}&\textbf{URI-Rs} & \textbf{URI-Ms}&\textbf{URI-Rs} & \textbf{URI-Ms}&\textbf{URI-Rs} & \textbf{URI-Ms}&\textbf{URI-Rs} \\ \hline

\href{http://web.archive.org}{\texttt{web.archive.org}}  & 34,819 & 9,968 & 36,385 & 10,075 & 39,092 & 10,265 & 40,258 & 10,353 \\

\href{https://archive.is}{\texttt{archive.is}}  & 2,330 & 1,289 & 2,562 & 1,359 & 3,093 & 1,479 & 3,229 & 1,526 \\

\href{http://archive-it.org}{\texttt{archive-it.org}} & 5,108 & 979 & 5,999 & 1,095 & 7,373 & 1,271 & 7,986 & 1,355 \\

\href{http://archive.bibalex.org}{\texttt{archive.bibalex.org}} & 4,169 & 675 & 4,764 & 762 & 5,827 & 891 & 6,176 & 941 \\

\href{http://webarchive.loc.gov}{\texttt{webarchive.loc.gov}} & 4,432 & 590 & 5,297 & 698 & 6,598 & 860 & 7,122 & 934 \\

\href{https://arquivo.pt}{\texttt{arquivo.pt}} & 1,762 & 456 & 2,158 & 553 & 2,935 & 689 & 3,154 & 758 \\

\href{https://www.webcitation.org}{\texttt{webcitation.org}} & 1,290 & 532 & 1,415 & 593 & 1,737 & 689 & 1,858 & 725 \\

\href{http://europarchive.org}{\texttt{europarchive.org}}  & 476 & 137 & \textbf{603} & \textbf{202} & 842 & 265 & 911 & 287 \\

\href{https://swap.stanford.edu}{\texttt{swap.stanford.edu}} & 651 & 145 & \textbf{797} & \textbf{200} & 1,072 & 261 & 1,176 & 283 \\

\href{https://vefsafn.is}{\texttt{vefsafn.is}} & 19 & 7 & 19 & 7 & \textbf{1,270} & \textbf{200} & 1,520 & 246 \\

\href{https://www.webharvest.gov}{\texttt{webharvest.gov}}  & \textbf{647} & \textbf{200} &  698 & 212 & 711 & 214 & 743 & 227 \\

\href{https://www.digar.ee/arhiiv}{\texttt{digar.ee}}  & 339 & 134 & 362 & 153 & \textbf{491} & \textbf{212} & 513 & 223 \\

\href{http://webarchive.proni.gov.uk/}{\texttt{webarchive.proni.gov.uk}} &144 & 52 & 156 & 58 & 224 & 84 & 316 & 141 \\

\href{https://perma.cc}{\texttt{perma.cc}}  & 58 & 38 & 74 & 50 & 76 & 52 & 101 & 71 \\

\href{http://www.collectionscanada.gc.ca}{\texttt{collectionscanada.gc.ca}} & 26 & 22 & 40 & 35 & 42 & 37 & 59 & 50 \\ 

\href{https://webarchive.org.uk}{\texttt{webarchive.org.uk}} & 12 & 5 & 12 & 5 & 12 & 5 & 27 & 8 \\

\href{http://nationalarchives.gov.uk}{\texttt{nationalarchives.gov.uk}} & 0 & 0 & 0 & 0 & 0 & 0 & 0 & 0 \\

\hline
\textbf{Total} & \textbf{56,282}  & \textbf{15,229} & \textbf{61,341}  & \textbf{16,057}  & \textbf{71,395}  & \textbf{17,474}  & \textbf{75,149}  & \textbf{18,128} \\
\end{tabular}
\vspace{-2mm}
\label{tab:archive-urim-count-all-method1}
\end{adjustwidth}
\end{table}

\begin{figure}
\centering
\begin{lstlisting}[frame=single,style=base]
python extract_urirs.py http://wayback.vefsafn.is/wayback/20041020191800id\_/http://www.w3.org/
http://www.csail.mit.edu/
http://www.google.com/
http://www.ilog.com/
http://www.inria.fr/
http://jigsaw.w3.org/css-validator/
http://www.w3.org/People/Raggett/tidy/
http://validator.w3.org/
http://www.w3.org/2004/MWeb/Overview.html
http://purl.org/rss/1.0/
...
\end{lstlisting}
\caption{An example of extracting URI-Rs from the HTML of the memento \href{http://wayback.vefsafn.is/wayback/20041020191800id_/http://www.w3.org/}{\texttt{wayback.vefsafn.is/wayback/20041020191800id\_/http://www.w3.org/}} (only 9 URI-Rs, out of 138, are shown). Notice that we used the option \texttt{id\_} in the URI-M to retrieve the archived unaltered, or raw, content of the memento.}
\label{script:extrcat-example}
\end{figure}

\begin{figure}[ht]
\centering
\begin{lstlisting}[frame=single,style=base]
% python timemap.py http://www.inria.fr/
19961230035541 https://$web.archive.org$/web/19961230035541/http://www4.inria.fr:80/
...
19961230035541 http://web.archive.$bibalex.org$:80/web/19961230035541/http://www4.inria.fr/
...
20140729051225 http://www.$webcitation.org$/6RQAbDGPm
20020808175122 http://$webarchive.loc.gov$/all/20020808175122/http://www.inria.fr/
20100731132417 http://wayback.$archive-it.org$/all/20100731132417/http://www.inria.fr/
...
19961230035541 http://$archive.is$/19961230035541/http://www.inria.fr/
...
19961013190926 https://$arquivo.pt$/wayback/19961013190926/http://www.inria.fr/
...
20110325131647 http://veebiarhiiv.$digar.ee$/a/20110325131647/http://www.inria.fr/
...
\end{lstlisting}
\caption{Downloading the TimeMap of the URI-R \href{http://www.inria.fr/}{\texttt{http://www.inria.fr/}}.}
\label{script:extrcat-example-timemap}
\end{figure}

With \textbf{Method 2}, we now have 200 URI-Rs for the following additional archives: 

\begin{itemize}
\item \href{https://www.webharvest.gov}{\texttt{webharvest.gov}}
\item \href{https://swap.stanford.edu}{\texttt{swap.stanford.edu}}
\item \href{https://vefsafn.is}{\texttt{vefsafn.is}}
\item \href{https://www.digar.ee/arhiiv}{\texttt{digar.ee}}
\item \href{http://europarchive.org}{\texttt{europarchive.org}}
\end{itemize}

{\raggedleft{}T}able \ref{tab:archive-urim-count-all} (column \textbf{Method 2}) shows the new archives that satisfy the condition of 200 URI-Rs and the four archives which still did not satisfy the condition. 


\subsection{Method 3: URI-Rs discovered in archives' published lists} \label{sec:publishedlists}

Some archives make lists of URI-Rs they collect available on the web. Archives may also publish lists of URI-Ms associated with each URI-R. We found these published collections for three archives (Table \ref{tab:publised-lists}) that had not met the 200 URI-R minimum.


\begin{table}
\setlength\tabcolsep{5pt}
\centering
\caption {\bf Archives' published lists of URI-Rs and URI-Ms.}

\begin{tabular}{llrr}
\hline
\textbf{Archive}& \textbf{List} & \textbf{URI-R}& \textbf{URI-M} \\
\hline

\shortstack[l]{\href{https://webarchive.org.uk}{\texttt{webarchive.org.uk}}\\~}  & \href{https://data.webarchive.org.uk/opendata/ukwa.ds.1/} {\shortstack[l]{\texttt{data.webarchive.org.uk/}\\\texttt{opendata/ukwa.ds.1/}}} & \shortstack[r]{26,910\\~} & \shortstack[r]{-\\~}  \\ \hline

\shortstack[l]{\href{http://www.collectionscanada.gc.ca}{\texttt{collectionscanada.gc.ca}}\\~\\~\\~}  & \href{http://collectionscanada.gc.ca/webarchives/url-list/index-e.html}{\shortstack[l]{
\texttt{collectionscanada.gc.ca}\\\texttt{/webarchives/url-list/i}\\\texttt{ndex-e.html}}} & \shortstack[r]{2,613\\~\\~\\~} & \shortstack[r]{27,232 \\~\\~\\~}\\ \hline

\shortstack[l]{\href{http://nationalarchives.gov.uk}{\texttt{nationalarchives.gov.uk}}\\~}  & \href{http://nationalarchives.gov.uk/webarchive/atoz/} {\shortstack[l]{\texttt{nationalarchives.gov.uk}\\\texttt{/webarchive/atoz/}}} & \shortstack[r]{4,956\\~} & \shortstack[r]{168,328\\~}  \\ \hline

\end{tabular}
\label{tab:publised-lists}
\end{table}

We downloaded the published list of URI-Rs only (URI-Ms were not included in this list) from the archive \href{https://webarchive.org.uk}{\texttt{webarchive.org.uk}}. Then, using the LANL Memento Aggregator we retrieved TimeMaps that at least contain one memento in the UK Web Archive, of the first 192 URI-Rs. Table \ref{tab:archive-urim-count-all} (column \textbf{Method 3}) shows that this method helps two archives to reach 200 URI-Rs (i.e., \href{http://webarchive.proni.gov.uk/}{\texttt{webarchive.pro ni.gov.uk}} and \href{https://webarchive.org.uk}{\texttt{webarchive.org.uk}}), but at the same time, a new web archive appears in the TimeMaps, \href{http://nationalarchives.gov.uk}{\texttt{nationalarchives.gov.u k}},
raising the total number of archives to 17. 


Next, we downloaded lists of URI-Rs and URI-Ms made available by the two web archives \href{http://www.collectionscanada.gc.ca}{\texttt{collectionscanada.gc.ca}} and \href{http://nationalarchives.gov.uk}{\texttt{nationalarchives.gov.uk}}.
 We only extracted the number required to reach 200 URI-Rs per archive. With this method, we did not need a Memento aggregator since the archives already provide a list of mementos, but for the sake of consistency, we used the LANL's Aggregator to download TimeMaps, so we can update information for the other archives. Table \ref{tab:archive-urim-count-all} (column \textbf{Method 3}) shows that for \href{https://perma.cc}{\texttt{perma.cc}} we only needed to discover 89 additional URI-Rs to reach 200 URI-Rs. Table \ref{tab:archive-urim-count-all-method2} shows how the number of URI-Rs/URI-Ms has increased after applying Method 3 to each of the three archives.

\begin{table}
\begin{adjustwidth}{-0.15in}{0in}
\centering
\scriptsize
\caption {\bf Applying Method 3 (using archives' published lists) for three archives.}
\begin{tabular}{l|rr|rr|rr}
\hline
&  \multicolumn{6}{c}{\textbf{Method 2}} \\ \cline{2-7}
 
&  \multicolumn{2}{c|}{  \href{https://webarchive.org.uk} {\textbf{\shortstack[l]{\\~\\webarchive.org.uk}}}}
&  \multicolumn{2}{c|}{  \href{http://www.collectionscanada.gc.ca} {\textbf{collectionscanada.gc.ca}}     }
&  \multicolumn{2}{c}{   \href{http://nationalarchives.gov.uk}{\textbf{nationalarchives.gov.uk}}}
 \\ 
 & & & & & & \\ \cline{2-7}

\textbf{Archive} & \textbf{URI-Ms}&\textbf{URI-Rs} & \textbf{URI-Ms}&\textbf{URI-Rs} & \textbf{URI-Ms}&\textbf{URI-Rs}\\ \hline

\href{http://web.archive.org}{\texttt{web.archive.org}} & 41,988 & 10,572 & 43,615  &  10,739 & 45,155  & 10,924 \\

\href{https://archive.is}{\texttt{archive.is}} &  3,370 &    1,590 & 3,426 & 1,619 & 3,471  &   1,654 \\

\href{http://archive-it.org}{\texttt{archive-it.org}} &  8,339 & 1,431 & 8,910   &  1,561 & 8,994  & 1,593 \\

\href{http://archive.bibalex.org}{\texttt{archive.bibalex.org}} &  6,630 & 1,009 & 7,043 & 1,091 & 7,286 & 1,148 \\

\href{http://webarchive.loc.gov}{\texttt{webarchive.loc.gov}} &  7,344 &  989 & 7,721  &   1,039 & 7,766  &   1,062  \\

\href{https://arquivo.pt}{\texttt{arquivo.pt}} &  3,298 &  819 & 3,379 & 855 & 3,430   &   895 \\

\href{https://www.webcitation.org}{\texttt{webcitation.org}} &  1,892     & 746 & 1,944 & 765 & 1,954 &  775  \\

\href{http://europarchive.org}{\texttt{europarchive.org}} &  963 &  311 & 986 & 320 & 992    &  324 \\

\href{https://swap.stanford.edu}{\texttt{swap.stanford.edu}} &  1,198 &   292 & 1,232  &    303 & 1,233  &    304 \\

\href{https://vefsafn.is}{\texttt{vefsafn.is}} &  1,626 & 270 & 1,703   &   289 & 1,715    &  294 \\

\href{https://www.webharvest.gov}{\texttt{webharvest.gov}} &  743  & 227 & 826 & 248 & 826  &    248  \\

\href{https://webarchive.org.uk}{\texttt{webarchive.org.uk}} &  \textbf{812} & \textbf{201} & 812 & 201 & 907 & 228 \\

\href{https://www.digar.ee/arhiiv}{\texttt{digar.ee}} &  515 &  225 & 518   &   228 & 518    &  228 \\

\href{http://webarchive.proni.gov.uk/}{\texttt{webarchive.proni.gov.uk}} &  \textbf{460} &  \textbf{201} & 462 & 203 & 480 & 213 \\

\href{http://nationalarchives.gov.uk}{\texttt{nationalarchives.gov.uk}} &  3    & 1  & 3     &   1 & \textbf{1011} & \textbf{200} \\

\href{http://www.collectionscanada.gc.ca}{\texttt{collectionscanada.gc.ca}} &  59 & 50 & \textbf{359} & \textbf{200} & 359 & 200 \\

\href{https://perma.cc}{\texttt{perma.cc}}  &  104 & 73 & 154   &   111 & 154  & 111  \\

\hline
\textbf{Total} &  \textbf{79,344} & \textbf{83,093}  & \textbf{19,773} & \textbf{19,007} & \textbf{86,251} & \textbf{20,401} \\
\end{tabular}
\label{tab:archive-urim-count-all-method2}
\end{adjustwidth}
\end{table}

\subsection{Method 4: Sending TimeMap requests directly to an archive } \label{sec:directtimeaps}

The LANL Memento Aggregator may serve cached TimeMaps \cite{bornand2016routing}, which may result in TimeMaps that do not contain recently created mementos. For this reason we decided to request TimeMaps for the already selected URI-Rs directly from \href{https://perma.cc}{\texttt{perma.cc}}. Figure \ref{script:timemap_perma} shows an example of  requesting the TimeMap of the URI-R \href{https://www.whitehouse.gov}{\texttt{www.whitehouse.gov}} from \href{https://perma.cc}{\texttt{perma.cc}} (the archive uses other domain names like \texttt{perma-archives.org}). It contains 57 mementos. By this method, we were able to obtain the additional 89 URI-Rs for \href{https://perma.cc}{\texttt{perma.cc}} shown in Table \ref{tab:archive-urim-count-all} (column \textbf{Method 4}).

\begin{figure}[ht]
\centering
\begin{lstlisting}[frame=single,style=base]
curl https://perma-archives.org/warc/$timemap$/*/$http://www.whitehouse.gov/$
<https://perma-archives.org/warc/timemap/*/http://www.whitehouse.gov/>; rel="self"; type="application/link-format"; from="Thu, 27 Aug 2015 17:14:18 GMT",
<http://www.whitehouse.gov/>; rel="original",
<https://perma-archives.org/warc/timegate/http://www.whitehouse.gov/>; rel="timegate",
<https://perma-archives.org/warc/20150827171418/http://www.whitehouse.gov/>; rel="memento"; datetime="Thu, 27 Aug 2015 17:14:18 GMT",
<https://perma-archives.org/warc/20150827171418/https://www.whitehouse.gov/>; rel="memento"; datetime="Thu, 27 Aug 2015 17:14:18 GMT",
<https://perma-archives.org/warc/20150831171426/https://www.whitehouse.gov/>; rel="memento"; datetime="Mon, 31 Aug 2015 17:14:26 GMT",
...
<https://perma-archives.org/warc/20180302185657/https://whitehouse.gov/>; rel="memento"; datetime="Fri, 02 Mar 2018 18:56:57 GMT",
<https://perma-archives.org/warc/20180302185657/https://www.whitehouse.gov/>; rel="memento"; datetime="Fri, 02 Mar 2018 18:56:57 GMT",
<https://perma-archives.org/warc/20180828214528/https://www.whitehouse.gov/>; rel="memento"; datetime="Tue, 28 Aug 2018 21:45:28 GMT
\end{lstlisting}
\caption{An example of requesting a TimeMap from Perma archive. The TimeMap of \href{https://www.whitehouse.gov}{\texttt{www.whitehouse.gov}} contains 57 mementos (only 6 are shown).}
\label{script:timemap_perma}
\end{figure}

\subsection{Filtering by download time, the maximum number of mementos, and HTTP status codes} \label{sec:filter-by-time}

At this point, the selected set contained 86,387 URI-Ms, 20,490 total, and 11,222 unique URI-Rs from 17 different web archives. For our target study, we downloaded the rewritten and raw mementos 10 times. We ran 17 parallel processes where each process downloaded mementos from a specific archive. We found that download time varies between web archives. For example, it took about 40 hours to download 733 mementos from \href{https://www.webharvest.gov}{\texttt{webharvest.gov}} and 12 hours to download 1,011 mementos from \href{http://nationalarchives.gov.uk}{\texttt{nationalarchives.gov.uk}}. Thus, we decided to change the number of mementos per archive to what could be downloaded within 40 hours, and the number of mementos must not exceed 1,600 per archive. This produced 18,472 mementos. Unfortunately, we did not check the HTTP status when selecting mementos to make sure they are ``200 OK'' or archival 4xx/5xx responses (i.e., they have the HTTP response header \texttt{Memento-Datetime} for the archives that support the Memento protocol). After selecting the 18,472 mementos, we found that about 10\%, 1,975, of these mementos had the HTTP status code of non-archival 4xx or 5xx (1,498 are from \href{http://archive.bibalex.org}{\texttt{archive.bibalex.org}}) as the example in Figure \ref{script:curl-503-ex} shows. Thus, we removed most of the 4xx/5xx mementos and kept only 130 (out of 1,975) because we wanted to keep track of these mementos. We could not replace those removed mementos because by this time we had already used the selected dataset in our study, and it was not possible to recover any excluded mementos. This resulted in 16,627 mementos remaining.

 \begin{figure}
\centering
\vspace{-2mm}
\begin{lstlisting}[frame=single,style=base]
% curl -I http://web.archive.bibalex.org/web/20051026134855/http:/www.anchorage.gc.ca/ 
HTTP/1.1 503 Service Unavailable
Server: Apache-Coyote/1.1
Set-Cookie: JSESSIONID=817B5F389A4566092459D914091A0961; Path=/; HttpOnly
Content-Type: text/html;charset=utf-8
Transfer-Encoding: chunked
Date: Sun, 31 Mar 2019 13:25:47 GMT
Connection: close
\end{lstlisting}
\vspace{-3mm}
\caption{Non archival HTTP 503 example. The HTTP response header \texttt{Memento-Datetime} is not included in the returned response.}
\label{script:curl-503-ex}
\vspace{-2mm}
\end{figure}

\subsection{Final set}

Table \ref{tab:archive-urim-count-final} shows the final numbers of selected URI-Rs and URI-Ms per archive (available on GitHub\footnote{\href{https://github.com/oduwsdl/mementos-fixity}{\texttt{https://github.com/oduwsdl/mementos-fixity}}}). The table shows that three archives have fewer than 200 URI-Rs for the following reasons: 
\begin{itemize}

\item{\href{https://perma.cc}{\texttt{perma.cc}}: It took about 40 hours to download 182 mementos from perma.cc, including the raw mementos.}

\item{\href{http://archive.bibalex.org}{\texttt{archive.bibalex. org}}: We removed 1,498 mementos because they returned the ``\texttt{503 Service Unavailable}'' HTTP response code.}


\item{\href{http://www.collectionscanada.gc.ca}{\texttt{collectionscanada.gc.ca}}: We removed mementos of two URI-Rs that returned the ``\texttt{503 Service Unavailable}'' HTTP response code.}

\end{itemize}

\begin{figure}
\centering
\setlength{\fboxsep}{0pt}%
\fbox{
\includegraphics[width=0.95\textwidth]{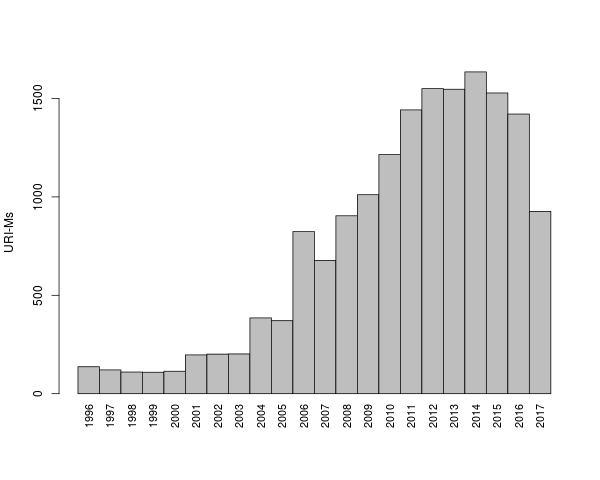}
}
\caption{URI-Ms per year. Note that we collected mementos in November 15, 2017. For this reason, the number of mementos from 2017 is less than the number of mementos in other years, 2010-2016 (i.e., no mementos with \texttt{Memento-Datetime} value after November 15, 2017).}
\label{img:urims_per_year}
\end{figure}

\begin{figure}
\centering
\setlength{\fboxsep}{0pt}%
\fbox{
\includegraphics[width=200pt]{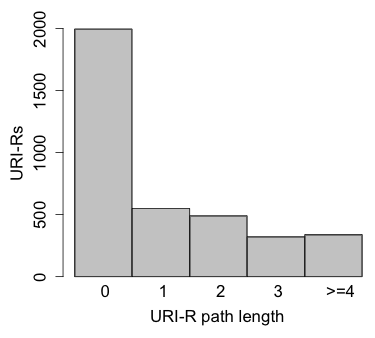}
}
\caption{URI-Rs per path length (54\% of URI-Rs are with zero path length).}
\label{img:urirs_per_paths}
\end{figure}

Figure \ref{img:urims_per_year} shows the distribution of URI-Ms between 1996 and 2017. The main reason for having fewer mementos in years 1996-2005 is because most web archives did not exist during those early years \cite{costa2017,Kim_Nowviskie_Graham_Quon_Alliance_20172}. 
Figure \ref{img:urirs_per_paths} shows the number of URI-Rs per path length. The number of distinct URI-Rs is 3,698, and of those 1,996 (54\%) have a path length of zero and the remaining 1,702 URIs (46\%) have a path length greater than or equal to one.

\begin{table}
\centering
\vspace{-2mm}
\caption {Final numbers in the selected set of  URI-Rs and URI-Ms.}
\begin{tabular}{lrr}
\textbf{Archive } & \textbf{URI-Rs}&\textbf{URI-Ms}\\ \hline
\href{http://web.archive.org}{\texttt{web.archive.org}} & 1,566 & 1,566 \\
\href{http://archive-it.org}{\texttt{archive-it.org}} & 1,338 & 1,383 \\
\href{https://archive.is}{\texttt{archive.is}} & 1,257 & 1,396 \\
\href{http://webarchive.loc.gov}{\texttt{webarchive.loc.gov}} & 1,059 & 1,594 \\
\href{https://arquivo.pt}{\texttt{arquivo.pt}} & 766 & 1,569 \\
\href{https://www.webcitation.org}{\texttt{webcitation.org}} & 720 & 1,585 \\
\href{http://europarchive.org}{\texttt{europarchive.org}} & 321 & 979 \\
\href{https://swap.stanford.edu}{\texttt{swap.stanford.edu}} & 302 & 1,222 \\
\href{https://vefsafn.is}{\texttt{vefsafn.is}} & 290 & 1,589 \\
\href{https://www.webharvest.gov}{\texttt{webharvest.gov}} & 247 & 712 \\ 
\href{https://www.digar.ee/arhiiv}{\texttt{digar.ee}} & 225 & 488 \\
\href{https://webarchive.org.uk}{\texttt{webarchive.org.uk}} & 221 & 349 \\
\href{http://webarchive.proni.gov.uk/}{\texttt{webarchive.proni.gov.uk}} & 209 & 469 \\
\href{http://nationalarchives.gov.uk}{\texttt{nationalarchives.gov.uk}} & 200 & 994 \\ 
\href{http://www.collectionscanada.gc.ca}{\texttt{collectionscanada.gc.ca}} & 198 & 351 \\ 
\href{https://perma.cc}{\texttt{perma.cc}} & 175 & 182 \\  
\href{http://archive.bibalex.org}{\texttt{archive.bibalex.org}} & 168 & 199 \\
\hline
\textbf{Total} & \textbf{9,262} & \textbf{16,627} \\
\end{tabular}
\label{tab:archive-urim-count-final}
\end{table}

\section{Conclusions}
In this paper we describe four methods to discover 16,627 mementos from 17 public web archives. 
We use the LANL Memento Aggregator to look up mementos by submitting the URI-R of original web pages (Method 1). For archives that have fewer than 200 URIs, we collect additional mementos by extracting URI-Rs from the HTML of already discovered mementos (Method 2). As our third method, we use published lists of original web pages and their associated mementos made available by several web archives. Finally, we request TimeMaps directly from the archive \href{https://perma.cc}{\texttt{perma.cc}} (Method 4). Even though the process of discovering mementos resulted in a total of 80,387 mementos (after applying the one memento per year filter), we downsampled this number to 16,627 due to our constraints of limiting to 1,600 URI-Ms per archive, being able to download all the mementos in less than 40 hours, and the condition that the number of URI-Rs per archive should be greater than or equal to 200.

\section{ACKNOWLEDGEMENTS}
\vspace{-2.8mm}
This work is supported in part by The Andrew W. Mellon Foundation (AMF) grant 11600663.

\bibliographystyle{splncs03}
\bibliography{discovering}

\end{document}